\documentclass[12pt]{iopart}
\usepackage{graphicx}
\usepackage{iopams}
\usepackage{setstack}

\begin{document}
\title[Dynamical stabilization]{Dynamical stabilization of  the bcc phase in lanthanum and thorium  by phonon-phonon interaction}
\author{P. Souvatzis$^{1}$, T. Bj\"orkman$^{2}$, O. Eriksson$^{2}$, P.Andersson$^{3}$, M. I. Katsnelson$^{4}$ and  S. P. Rudin$^{1}$}
\address{$^{1}$ Theoretical Division, Los Alamos National Laboratory, Los Alamos, New Mexico 87545, USA}
\address{$^{2}$ Department of Physics, Uppsala University,
Box 530, SE-75121, Uppsala, Sweden}
\address{$^{3}$ FOI, Swedish Defence  Research Agency, SE-164 90 Stockholm, Sweden }
\address{$^{4}$ Institute for Molecules and Materials, Radboud
University Nijmegen, NL-6525 ED, Nijmegen, The Netherlands}
\ead{petros.souvatzis@gmail.com}
\date{\today }
\begin{abstract}
The recently developed self consistent {\it ab initio} lattice
dynamical method (SCAILD) has been applied to the high temperature bcc
phase of La and Th which are dynamically unstable at low temperatures. The bcc phase of these metals is found to be stabilized by phonon-phonon interactions. The calculated high
temperature phonon frequencies for La are found to be in good
agreement with the corresponding experimental data.
\end{abstract}

\maketitle

\section{Introduction}

The unique physical and chemical properties of the actinides and
lanthanides have attracted interest for decades. Several reviews
have been written on the subject (see, e.g.
Refs.\cite{borje1,olle1,handbook}), identifying different
properties which make these two series of the Periodic Table
unique. The uniqueness stems from the progressive filling  of
the f-shell  as one traverses each series. In
general one  observes that the 4{\it f}-electrons in the lanthanides
behave in a localized way, much in the same way they do in the
atomic configurations. For the actinides (where the 5{\it f} shell becomes 
filled) a more complex behavior is observed:  for the latter
part of the series (from Am and onward) the 5{\it f} states are
localized, whereas for the earlier part of the series (from Th to
Np) the 5{\it f} states are itinerant and form energy bands. This
becomes evident when, e.g., considering the cohesive energy and
equilibrium volumes\cite{borje2} as well as the structural
properties\cite{per1,per2}. The element Pu sits right on the
border between these parts of the actinide series, demonstrating
transition from itinerant {\it f}-electrons in $\alpha$-phase to the
localized one in $\delta$-phase\cite{Pu1,Pu2}, and hence the
electronic structure and many other properties of this material are
extremely complex.

An experimental fact which is common for the lanthanides and
actinides is that these elements do not usually crystallize in the
body centered cubic (bcc) structure at low enough temperatures, but they melt frequently
out of this structure. In this paper we have chosen to investigate
this complex behavior, by studies of representative materials of
the two series, i.e. La and Th. At low temperatures La and Th have
the double hexagonal closed packed (dhcp) and face centered cubic (fcc) structure, respectively. The low-temperature
structural properties are perfectly understood from first
principles theory \cite{skriver,john}, showing that it is the
filling of the d-band that is responsible for the structure of
La\cite{skriver}, whereas  Th is an intermediate system which has
been shown to have its crystal structure governed by a balance of
the occupation of the 5{\it f} and the 6{\it d} states\cite{borje3}. Hence the
two elements chosen in this study represent different structural
and electronic properties in the ground state phases. Despite
these marked differences these materials exhibit a typical
behavior of the 4{\it f} and 5{\it f} series, namely that they melt out of the
bcc crystal structure. It is the purpose of this report to
show that  the interaction between different phonon modes is responsible for the 
dynamical stabilization of  the  high temperature phase of
these two series.

\section{Details of the calculations}
The lattice dynamical calculations  of  La and Th at elevated
temperatures  were performed using the recently developed 
self-consistent {\it ab initio} lattice dynamical method
(SCAILD)\cite{petros1}. In the SCAILD method one displaces the atoms of a
supercell along several phonon modes simultaneously, in order to include the effects of
phonon-phonon interactions upon the phonon frequencies at elevated
temperatures. The
simultaneous presence of several frozen phonons in the supercell 
introduces geometric disorder, i.e., entropy, which in turn
re-normalizes the phonon frequencies. In the SCAILD approach, the
atomic displacements $\mathbf{U}_{\mathbf{R}}$ of the atoms
located at equilibrium Bravais lattice sites $\mathbf{R}$, are
taken as superpositions of all the phonon modes $s$,  with wave vectors $\mathbf{q}$ found to be
commensurate with the cell, i.e, 
\begin{equation}
\mathbf{U}_{\mathbf{R}}
= \frac{1}{\sqrt{N}}\sum_{\mathbf{q},s}\mathcal{A}_{\mathbf{q}s}
\mathbf{\epsilon}_{\mathbf{q}s}e^{i\mathbf{q}\mathbf{R}}.\label{eq:SUPERPOS}
\end{equation}
Here $\mathbf{\epsilon}_{\mathbf{q}s}$ are the eigenvectors of the dynamical matrix
\begin{equation}\label{eq:DYNMAT}
\mathcal{D}(\mathbf{q})=\frac{1}{M}\sum_{\mathbf{R}}\Phi(\mathbf{R})e^{-i\mathbf{q}\mathbf{R}},
\end{equation}
where $\Phi(\mathbf{R})$ are the force constant matrices and
\begin{eqnarray}
\mathcal{A}_{\mathbf{q}s}= \pm \sqrt{ \frac{\langle \mathcal{Q}_{\mathbf{q}s}\mathcal{Q}_{-\mathbf{q}s}\rangle}{M}}=\pm \sqrt{
\frac{\hbar}{M\omega_{\mathbf{q}s}}\Big [ \frac{1}{2} + n \Big
(\frac{\hbar\omega_{\mathbf{q}s}}{k_{B}T} \Big ) \Big ]}, 
\label{eq:AMPL}
\end{eqnarray}
where $n(x) = 1/( e^{x}-1)$ is the Planck function, $M$ is the
mass of atoms,  and $\mathcal{Q}_{\mathbf{q}s}$ are the canonical
phonon operators  appearing together with the canonical phonon
momentum, $\mathcal{P}_{\mathbf{q}s}$, in the harmonic
Hamiltonian,
$\mathcal{H}_{h}=\sum_{\mathbf{q},s}\frac{1}{2}(\mathcal{P}_{\mathbf{q}s}\mathcal{P}_{-\mathbf{q}s}+\omega_{\mathbf{q}s}^{2}
\mathcal{Q}_{\mathbf{q}s}\mathcal{Q}_{-\mathbf{q}s})$.

The SCAILD method alternates between setting up atomic
displacements based on phonon frequencies and evaluating the
phonon frequencies from {\it ab initio} calculated forces acting
on the displaced atoms. For the first iterative step, the forces
stem from a direct force method calculation, see, e.g.,
Ref.\cite{DF1,PETROS}. The phonon frequencies and eigen
vectors corresponding to commensurate wave vectors $\mathbf{q}$, serve to calculate
a set of atomic displacements $\mathbf{U}_{\mathbf{R}}$ through
Eqn.(\ref{eq:SUPERPOS}) and (\ref{eq:AMPL}). A first principles
calculation provides the Hellman-Feynman forces acting on the
displaced atoms, and a new set of phonon frequencies  are obtained
from the Fourier transform $\mathbf{F}_{\mathbf{q}}$ of the
forces,
\begin{equation}\label{eq:FOURIER}
\bar{\omega}_{\mathbf{q}s} =
\Big [ \frac{\mathbf{\epsilon}_{\mathbf{q}s} \cdot \mathbf{F}_{\mathbf{q}}}{\mathcal{A}_{\mathbf{q}s}M}
\Big ]^{1/2}.
\end{equation}
To clarify how  the interactions between different phonon modes affect 
the renormalization of the phonons, we here also give the alternative expression for the new set of phonon frequencies  $\bar{\omega}_{\mathbf{q}s}$ in terms of  anharmonic potential coefficients  \cite{petros2}
\begin{equation} \label{eq:E16}
\bar{\omega}_{\mathbf{q}s}^{2}= \omega_{\mathbf{q}s}^{2}\Big (1+  
\frac{\sqrt{M}}{2}\frac{\mathcal{A}_{\mathbf{q}s}}{\omega_{\mathbf{q}s}^{2}}\sum_{\mathbf{q}_{1},\mathbf{q}_{2}}\sum_{s_{1},s_{2}}\mathcal{R}(\mathbf{q},\mathbf{q}_{1},\mathbf{q}_{2},s,s_{1},s_{2})
\frac{\mathcal{A}_{\mathbf{q}_{1}s_{1}}\mathcal{A}_{\mathbf{q}_{2}s_{2}}}{\mathcal{A}_{\mathbf{q}s}^{2}} +\cdots \Big ). 
\end{equation}
Here
\begin{eqnarray}\label{eq:E14}
\mathcal{R}(\mathbf{q},\mathbf{q}_{1},\mathbf{q}_{2},s,s_{1},s_{2})=  \nonumber \\
\frac{1}{(MN)^{3/2}}\sum_{\mathbf{R},\mathbf{R}_{1},\mathbf{R}_{2}}\sum_{\alpha,\beta,\gamma}
\Phi_{\alpha\beta\gamma}(\mathbf{R},\mathbf{R}_{1},\mathbf{R}_{2})\mathbf{\epsilon}_{\mathbf{q}s\alpha}\mathbf{\epsilon}_{\mathbf{q}_{1}s_{1}\beta}
\mathbf{\epsilon}_{\mathbf{q}_{2}s_{2}\gamma}e^{i(\mathbf{Rq}+\mathbf{R}_{1}\mathbf{q}_{1}+\mathbf{R}_{2}\mathbf{q}_{2})} , \\ \nonumber
\end{eqnarray}
where $\Phi_{\alpha\beta\gamma}(\mathbf{R},\mathbf{R}_{1},\mathbf{R}_{2})$ are the third order 
coefficients appearing in the anharmonic part of the lattice dynamical hamiltonian, see for instance  Ref. \cite{WALLE, petros2}. These  coefficients  describe,  in terms of anharmonic changes to the nuclear potential energy,  the underlying electronic structure response to the atomic displacements.  Even though only third order anharmonic terms are explicitly written out in Eqn. (\ref{eq:E16}), it is now apparent 
that the simultaneous excitations of several  commensurate phonons  in the SCAILD calculation, renormalizes the phonon frequencies
through anharmonic terms such as $\mathcal{R}(\mathbf{q},\mathbf{q}_{1},\mathbf{q}_{2},s,s_{1},s_{2})$. Thus these anharmonic terms can be viewed as the mediators  of the phonon-phonon interaction.

The symmetry of the frequencies obtained through  Eqn. (\ref{eq:FOURIER}) is restored through 
\begin{equation}\label{eq:symm}
\Omega_{\mathbf{q}s}^{2}= \frac{1}{m_{\mathbf{q}}}\sum_{\mathcal{S}\in \mathcal{S}(\mathbf{q})}\bar{\omega}_{\mathcal{S}^{-1}\mathbf{q}s}^{2},
\end{equation}
where $\mathcal{S}(\mathbf{q})$ is the symmetry group of the wave
vector $\mathbf{q}$ and $m_{q}$ the number of elements of the
group. From these symmetry restored frequencies a new set of frequencies 
\begin{equation}\label{eq:mean}
\omega_{\mathbf{q}s}^{2}(N_{I})= \frac{1}{N_{I}}\sum_{i=1}^{N_{I}}\Omega_{\mathbf{q}s}^{2}(i),
\end{equation}
are calculated. Here  $\Omega_{\mathbf{q}s}(i)$, $i=1,...,N_{I}$ are the symmetry
restored frequencies from the $N_{I}$ previous iterations. The frequencies obtained through Eqn. (\ref{eq:mean}) are then used to calculate a new set of atomic displacements
$\mathbf{U}_{\mathbf{R}}$, through Eqn. (\ref{eq:SUPERPOS}) and
(\ref{eq:AMPL}), which in turn serve to calculate a new set of
forces. The iterative loop is terminated when the frequencies obtained through Eqn. (\ref{eq:mean}) have converged. For a more detailed discussion of the SCAILD method we
refer the reader to the work presented in Ref.
\cite{petros2, plutten}.

From the converged commensurate phonon frequencies and their corresponding 
eigen vectors, the force constant matrices $\Phi(\mathbf{R})$ of La and Th were calculated 
by inverse Fourier transformation. The force constant matrices were then used to calculate 
the dynamical matrices  $\mathcal{D}(\mathbf{q})$ on a  mesh of 22776  q-points in the irreducible Brillouin zone (100x100x100 Monkhorst-Pack mesh). By diagonalization of  the dynamical matrices on this mesh and by applying a gaussian smearing of 0.05 THz to  each phonon frequency, the phonon density of states (DOS) was obtained for bcc-La and bcc-Th.

Regarding the details of the force calculation we
used the VASP package \cite{VASP}, within the generalized gradient
approximation (GGA). The PAW potentials used required energy
cutoffs of  151 eV for La and  200 eV for Th. Monkhorst-Pack grids
with a 6x6x6  mesh were used together with 0.2 eV
Methfessel-Paxton smearing in all the calculations. The supercells
used were all 64 atom cells,  obtained by increasing the bcc
primitive cell 4 times along the 3 primitive lattice vectors.
%
%
\begin{figure}[tbp]
\begin{center}
\includegraphics*[angle=0,scale=0.7]{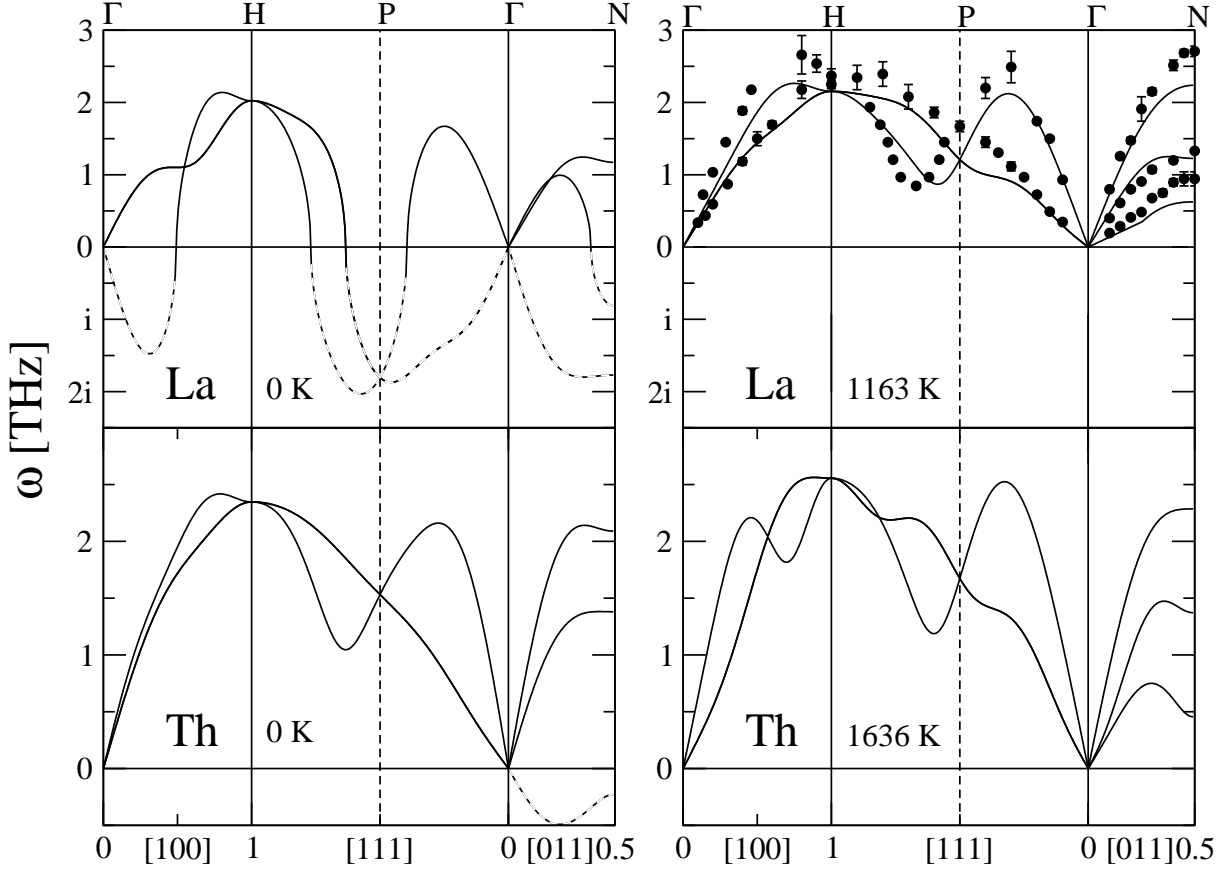}
\caption{The right hand panel, finite temperature SCAILD calculation of La and Th at 1163 K and 1636 K respectively. The left hand panel, standard direct force method
calculation \cite{DF1} of La and Th. Here the dashed curves in the left hand panel indicate imaginary frequencies. The filled circles and error bars  are the experimental data for La of Ref. \cite{LaEx} measured at 1163 K.}
\label{fig:ACT}
\end{center}
\end{figure}
\begin{figure}[tbp]
\begin{center}
\includegraphics*[angle=0,scale=0.7]{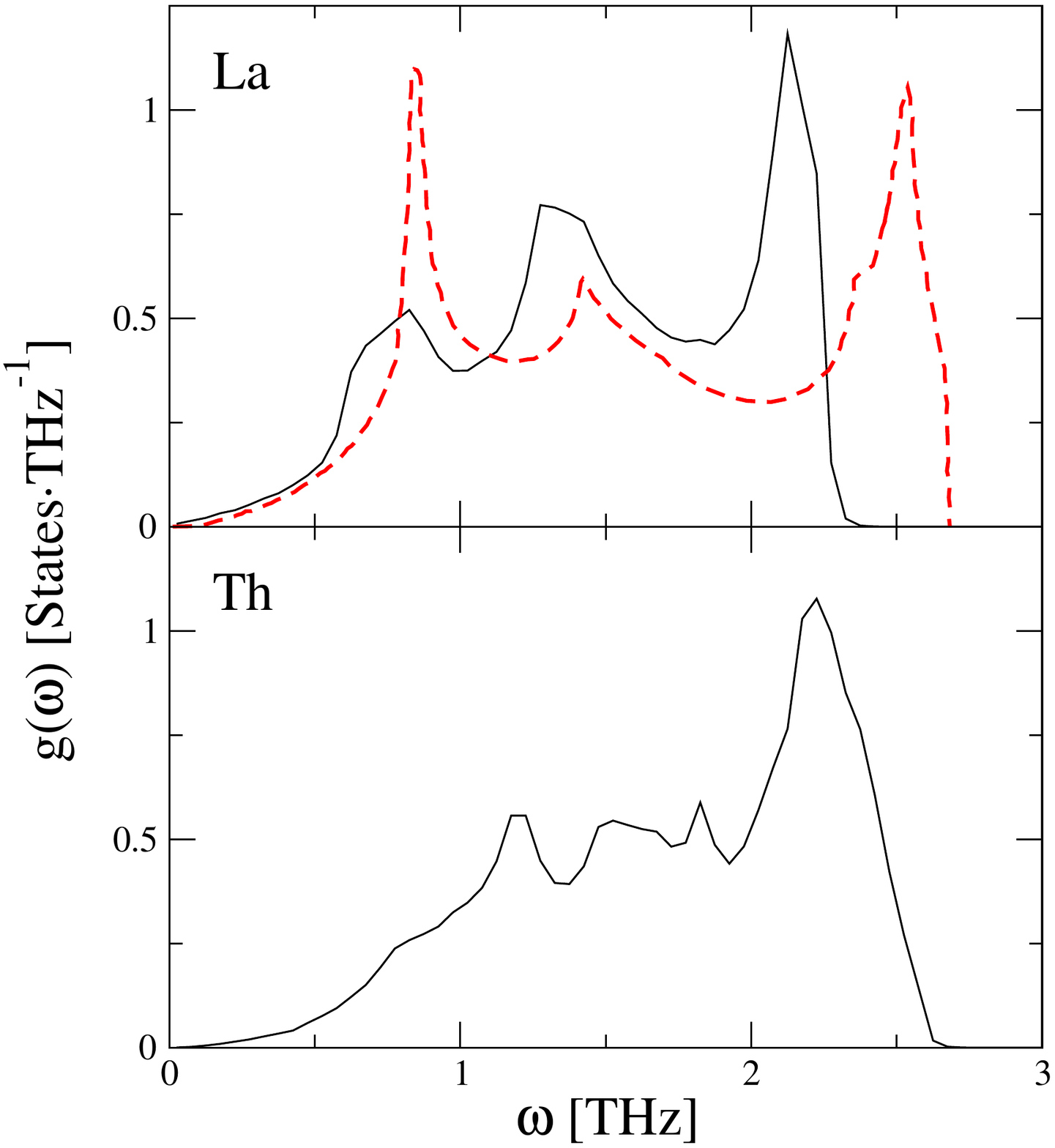}
\caption{(color online)The phonon density of states for bcc La and  bcc Th. The black curves are  the  1163 K and 1636 K  SCAILD calculations for La and Th, respectively.
 The dashed red line is the experimental bcc La data calculated from the 1163 K measurements of G\"uthoff {\it et al} \cite{LaEx}.  Both the experimental and the theoretical density of states $g(\omega)$ have been normalized so that $\int_{0}^{\infty}g(\omega)d\omega = 1$.}
\label{fig:DOS}
\end{center}
\end{figure}

In order to get a
quantitative measure of the quality of the PAW potentials used,
the equilibrium lattice constants and bulk moduli were calculated
and compared to their respective experimental values. Table
\ref{tab:table0} shows the results of these calculations together
with experimental data. From these results no immediate doubt can
be raised concerning the electronic structure part of the La
and Th calculations, since calculated equilibrium properties are
in a good agreement with experimental observations.
\begin{table}
\begin{center}
\begin{tabular}{l |  r @{.} l c r @{} l c r @{.} l c r @{} l c r @{} l  }
&
\multicolumn{2}{c}{$V_{0}^{(theor)}$ } & &
\multicolumn{2}{c}{$B^{(theor)}$ } & &
\multicolumn{2}{c}{$B'^{(theor)}$ } & &
\multicolumn{2}{c}{$V_{0}^{(exp)}$ } & &
\multicolumn{2}{c}{$B^{(exp)}$ } \\
\hline
La  & 37&3 & &  26&  & & 2&16  & & 37.7$^{a}$&   & & 27.9$^{a}$&  \\
Th  & 32&1 & &  58&  & & 2&53  & & 32.9$^{b}$&   & & 57.7$^{c}$& \\
\end{tabular}
\caption[]{ The zero temperature theoretical equilibrium volumes $V_{0}$
({\AA}$^{3}$), bulk moduli $B$ (GPa) and pressure derivatives of
the bulk moduli $B'=\partial B/ \partial P$ of the metals La and
Th, here presented together with experimental room temperature data. The structures
corresponding to the experimental data are hcp and fcc  for La and Th, respectevely.
$^{a}$Ref. \cite{VLa}, $^{b}$Ref. \cite{BLa},
$^{c}$Ref. \cite{VOLBTh}} \label{tab:table0}
\end{center}
\end{table}
%
\section{Results}
Figure \ref{fig:ACT} shows the calculated phonon dispersions for
the bcc phases of La and Th at temperatures 1163 K and 1636 K,
respectively. In the case of La, the corresponding
experimental data of Stassis {\it et al} \cite{LaEx} is displayed.
Furthermore, we also observe that the present  zero temperature calculations of La, reveal the same dynamical 
instabilities as in the previous work of Persson {\it et al} \cite{Persson}.
The finite temperature calculations predict the stability of the
bcc phase of both La and Th  by promoting the frequencies of the
phonons along the $\Gamma$ to $N$ and $\Gamma$ to $H$ symmetry
lines and around the $P$ symmetry point, from imaginary to
real. The finite temperature calculations of  the bcc-La phonons
result in an overall quantitative agreement with experimental
values. Smaller deviations are observed around the H, P and N point
of the Brillouin-zone, most likely due to finite size effects of
the supercell used in the calculations. The frequencies at commensurate wave vectors are affected by the
limited cell size due to the dependence on the number of  interacting phonons (the calculation only takes into account  interactions between  phonons with commensurate wave vectors). This differs from the direct force method, where 
the cell size affects the interpolation of phonon frequencies between commensurate wave vectors, but not the frequencies at 
commensurate wave vectors.
For an example  of size effects in SCAILD calculations see Ref. \cite{plutten}. From Eqn. (\ref{eq:AMPL}) and (\ref{eq:E16}) it can be seen that, at high enough temperatures, the relative shift of the squared frequency is to  first order inversely proportional to the harmonic frequency: $\delta \omega_{\mathbf{q}s}^{2}/\omega_{\mathbf{q}s}^{2}\sim \omega_{\mathbf{q}s}^{-1}$. Hence the transverse phonon modes, with their generally smaller phonon frequencies (compared to the longitudinal modes) are more strongly affected by the phonon-phononon interaction.

Figure \ref{fig:DOS} shows the calculated DOS of bcc-La and bcc-Th at 1163 K and 1636 K, respectively,
together with the experimental 1163 K data for La of G\"uthoff
{\it et al} \cite{LaEx}. Here the most obvious discrepancy is that
the experimental spectrum is $\sim0.5$ THz broader than the
theoretical one. Also, compared with the theoretical DOS of La the
corresponding experimental DOS  has considerable more weight
located at $\sim0.8$ THz. This lack of agreement between theory
and experiment is however not surprising, since the electronic
part of the bcc-La calculation has  been performed without
spin-orbit coupling (only scalar relativistic). 
%

Application of the SCAILD method in the calculation to the lattice
dynamical properties of the rare earths La and Th, shows that the
bcc structure of these elements is dynamically stabilized by
phonon-phonon interactions. 

\ack
The Department of Energy supported this work under Contract
No.~DE-AC52-06NA25396. 

\section*{References}
\begin{thebibliography}{22}
\bibitem{borje1}Johansson B and Li S 2007 J. of Alloys and Compounds {\bf 444} 202
\bibitem{olle1}Wills J M and Eriksson O 2000 ``Ground state propertiess of the actinide elements: a theoretical overview,
past, present and future'', LA-Science vol 26  128
\bibitem{handbook}See articles in "Handbook on the Physics and Chemistry of the actinides" edited by A.J.Freeman and G.H Lander (North Holland, Amsterdam, 1984).
\bibitem{borje2}Johansson B and Rosengren A 1975  Phys. Rev B {\bf 11} 1367
\bibitem{per1} S\"oderlind P, Eriksson O,  Johansson B , Wills J M , and Boring A M 1995 Nature {\bf 374} 524
\bibitem{per2} S\"oderlind P , Wills J M, Johansson B and Eriksson O 1997 Phys. Rev. B {\bf 55} 1997.
\bibitem{Pu1} Katsnelson M I ,  Solovyev I V  and Trefilov A V 1992
JETP Lett. {\bf 56} 572
\bibitem{Pu2} Savrasov S Y ,  Kotliar G  and Abrahams E 2001 Nature 
{\bf 410} 793
\bibitem{skriver} Skriver H L 1985 Phys. rev. B {\bf 31} 1909
\bibitem{john} Wills J M and Eriksson O 1992 Phys. Rev. B {\bf 45} 13879
\bibitem{borje3} Johansson B,  Ahuja R , Eriksson O and Wills J M 1995  Phys. Rev. Lett. {\bf  75} 280
\bibitem{petros1}Souvatzis P, Eriksson O, Katsnelson M I and Rudin S P 2008 Phys. Rev. Lett. {\bf 100} 095901
\bibitem{DF1}Kunc K , and Martin R M 1982 Phys. Rev. Lett {\bf 48}  406
\bibitem{PETROS}Souvatzis P,  Delin A, and Eriksson O 2006 Phys. Rev. B {\bf 73} 054110-054116
\bibitem{WALLE} Wallace D C 1972 {\it Thermodynamic of Crystals} (Dover  New York)
\bibitem{petros2}Souvatzis P , Eriksson O , Katsnelson M I  and Rudin S P 2009  Comput. Matter. Sci. {\bf 44} 888-894
\bibitem{plutten} Souvatzis  P and  Rudin S P 2008 Phys. Rev. B {\bf 78} 184304
\bibitem{VASP} Kresse  G \& Furthmuller J 1996 Phys. Rev. B {\bf 54} 11169
\bibitem{LaEx}G\"uthoff F, Petry W, Stassis C, Heiming A, Hennion B , Herzig C and Trampenau J 1993 Phys. Rev. B {\bf 47} 2563
\bibitem{Persson} Persson K, Ekman M and Ozoli\c{n}\v{s} V 2000 Phys. Rev. B {\bf 61} 11221
\bibitem{VLa} {\it Chemical Rubber Company handbook of chemistry and physics} 89th edition 2008 (CRC Press)
\bibitem{BLa}Benedict U 1987 J. Less Common Met. {\bf 128} 7-45
\bibitem{VOLBTh} Armstrong P E, Carlson O N and Smith J F 1959 J. Appl. Phys. {\bf 30} 36
\endbib


\end{document}